# Electrical and structural properties of MgB$_2$ films prepared by sequential deposition of B and Mg on the NbN buffered Si(100) substrate

Š. Chromik, Š.Gaži, V. Štrbík, M. Španková, I. Vávra, L. Satrapinsky and Š.Beňačka

*Institute of Electrical Engineering, Slovak Academy of Sciences, Dúbravská cesta 9, 841 04 Bratislava, Slovak Republic*

C.J. van der Beek

*Laboratoire des Solides Irradiés, CNRS. UMR 7642, Ecole Polytechnique, 91128 Paiseau, France*

P. Gierlowski

*Institute of Physics, Polish Academy of Sciences, Al. Lotnikow 32/46, 02 668 Warsaw, Poland*

**Abstract**

We introduce a simple method of an MgB$_2$ film preparation using sequential electron-beam evaporation of B-Mg two-layer (followed by in-situ annealing) on the NbN buffered Si(100) substrate. The Transmission Electron Microscopy analyses confirm a growth of homogeneous nanogranular MgB$_2$ films without the presence of crystalline MgO. A sensitive measurement of temperature dependence of microwave losses shows a presence of intergranular weak links close the superconducting transition only. The MgB$_2$ films obtained, about 200 nm thick, exhibit a maximum zero resistance critical temperature of 36 K and critical current density of $3\times10^7$ A/cm$^2$ at 13.2 K.

Corresponding author: Dr. Štefan Chromik

Institute of Electrical Engineering

Slovak Academy of Sciences

Dúbravská cesta 9, 841 04 Bratislava, Slovak Republic

Tel.: +421-2-54775826, ext. 2339,

Fax.: +421-2-54775816

E-mail address: elekchro@savba.sk

Recently we have published the properties of $MgB_2$ thin films prepared by sequential deposition of a multilayer consisting of B and Mg[1-3] followed by in-situ annealing in Ar atmosphere. A detailed study of these films has shown that in spite of a strong interdiffusion of B and Mg during in-situ annealing some maxima and minima of these elements still exist in the final films for a given thickness of Mg and B in the multilayer. Rutheford Backscattering Spectroscopy (RBS) analysis has revealed[2] a maxima of the $O_2$ between the Mg and the following B films in the concentration profiles of the as-deposited Mg-B multilayer where the maximum of the $O_2$ is appearently a consequence of the interaction of the Mg film with the surrounding atmosphere (vacuum ~ $10^{-4}$ Pa) before the deposition of the next B film. The presence of the $O_2$ in the $MgB_2$ film is usually linked with the creation of the MgO particles which may negatively influence superconducting properties.[4,5] The Si substrates which are the most attractive substrates for the applications of the $MgB_2$ films offer an $MgB_2$ film with zero resistance critical temperature below $T_{C0} \approx 27$ K.[1,3,4] We have observed an interaction of the Si substrate with the MgB$_2$ in the form of the Mg$_2$Si.[6]

This letter presents a process of the MgB$_2$ superconducting film preparation which avoids the most of the obstacles mentioned above. First, the Si(100) substrate was covered by a 0.1 μm thick NbN buffer layer grown by reactive magnetron sputtering in a mixture of N and Ar at room temperature to limit the interdiffusion between the substrate and the MgB$_2$ film. Second, we decided for the sequential deposition of the B-Mg two-layer, where the thickness of the B and Mg were 80 nm and 240 nm, respectively (100% excess of the Mg comparing to the stoichiometric composition). The advantage is only one B-Mg interface in the as-deposited film. In this case the Mg at the interface B-Mg is not exposed to the vacuum background during the period necessary for electron-beam evaporation of the next B film as it is in the case of the multilayer film. We suppose the diffusion of the Mg into the bottom B film during the in-situ annealing (post-deposition) process in Ar atmosphere. Briefly, the as-deposited films were in-situ heated at a rate of 10 °C / min to 280 °C and held there for 30 min at an Ar partial pressure of 0.06 Pa. Subsequently, the pressure was increased up to 16 Pa, the temperature was increased at a rate of 40 °C / min to 700 °C and kept there for 10 min. Finally, the Ar pressure was raised to $10^3$ Pa and the samples were cooled to room temperature.

Transmission Electron Microscopy (TEM) (JEOL 1200EX) and X-ray analyses of the prepared NbN buffer layers have shown the presence of an unreacted Nb in the polycrystalline NbN film. In-situ pre-annealing of the substrate with the NbN layer at the temperature of 730 °C and the pressure of $10^{-4}$ Pa for 15 min was applied to improve the composition and crystallinity of the NbN layer before the deposition of the MgB$_2$ film.

This step avoids the presence of the unreacted Nb. In the Selected Area Diffraction (SAD) pattern diffraction rings belonging to hexagonal and cubic NbN structure are present (Fig.1).

The superconducting properties of the $MgB_2$ films were characterized by resistance measurements (standard dc four probe technique). The maximum zero resistance critical temperature of $T_{C0} \approx 36$ K was achieved. The detailed TEM study reveals nanogranular character of the film with a maximum grain size about 10 nm. We detected the same density of the randomly oriented grains near the $MgB_2$/NbN interface as it was observed in the bulk of the $MgB_2$ film which is an evidence of the film homogeneity. The SAD pattern consists of concentric rings that belong to the hexagonal $MgB_2$ phase. In addition, the other diffraction rings correspond to hexagonal NbN phase only (Fig.2). This is a difference comparing to the mixture of hexagonal and cubic NbN phases after pre-annealing step. It can be a consequence of the additional annealing of the NbN during in-situ thermal treatment of the as-deposited B-Mg double layer. No rings, unlike to other authors,[4,5] belonging to the MgO particles were observed.

To receive more complex information about the bulk properties of the films a method of temperature dependence of microwave losses was applied. We use $TE_{011}$ circular Copper resonator filled by sapphire rod with a 8.9 GHz resonant frequency and quality factor of $Q_0 \cong 2000$ without the sample. Microwave losses caused by the presence of a weak link medium and dynamics of vortices were analyzed using magnetic modulation microwave absorption (MMMA) mode. Microwave losses measured by MMMA showed the presence of weak link-type losses at the transition of the sample into the superconducting state only (Fig.3, $B_{mod} = 0.05$ mT). This is in contrary to our previous results obtained in [1,4] where weak link losses were observed in large temperature range below $T_{C0}$ too. At small modulation amplitude and lower temperatures too, only viscous vortex motion type losses are present (Fig.3, $B_{mod}$ = 0.02 mT).[7] Both these results – no MgO observed by TEM in the bulk of the film and no presence of weak links below $T_{C0}$ - are in a good agreement with the high value of the critical current density $J_C \approx 3 \times 10^7 \, A/cm^2$ at 13.2 K. $J_C$ was determined using the magneto-optical study (Fig4a,b). $J_C$ vs B was derived from the hysteresis in the magnetic induction measured in the area (denoted by the white rectangle) in the series of such images, both in increasing and decreasing magnetic field. Fig.4b shows almost field independent $J_C$ at temperature 13.2 K in the field range 0-300 Gauss. The observed value of $J_C$ is comparable with the values published by Zheng et al.[8] on the epitaxial $MgB_2$ films prepared on the monocrystalline 6H-SiC substrate. This supports the idea about the transparency of the grain boundaries to current flow.[9] The similar value of the $J_C$ is presented in the work devoted to the magnetic properties of our $MgB_2$ films grown on the NbN buffered Si substrate.[10]

Usually the $MgB_2$ films were partially covered with Mg or MgO particles originated from the excess of Mg. However, in some cases, it was possible to apply point contact spectroscopy. The dependence of differential conductance vs. voltage was measured on normal metal (N)-superconductor (S) point contact. The best results were obtained when mechanically sharpened gold wire has been used as a tip. A typical characteristic (Fig.5) shows that Andreev reflection is a dominant component of electrical current through N-S boundary. Such behavior can be described by Blonder-Tinkham-Klapwijk (BTK) model.[11] The results (open circles in Fig.5) show two gap structure which is in a good correspondence with theoretical two-gap scenario in the clean limit.[12] Fitted values were $\Delta_1$ = 2 meV, $\Delta_2$ = 7.2 meV, $\Gamma$ = 0.01 and Z = 0 (where $\Gamma$ is smearing parameter and Z is barrier strength) for both gap. The fact that maximum differential conductance reaches the value only 1.2 can be an evidence of few channels for N-S boundaries. Together with the presence of clean limit, these results are in a good agreement with the TEM and microwave observations.

In conclusion, NbN buffered Si(100) substrates were used for the preparation of the $MgB_2$ films grown by sequential evaporation of the B and Mg. We have shown that the application of the NbN buffer layer, deposited at room temperature and pre-annealed in the in-situ step, before the process of the $MgB_2$ film preparation, results in improved superconductivity and transport properties of the final films. In spite of nanogranular character of the films, the critical current density $J_C \approx 30 \times 10^7$ A/cm$^2$ at 13.2 K (determined using the magneto-optical study) is comparable to the values for epitaxial $MgB_2$ films. To the best of our knowledge, the maximum zero resistance critical temperature 36 K is the highest value for in-situ prepared polycrystalline $MgB_2$ films on the Si substrate until now.

This work was supported by Slovak Grant Agency for Science (Nos 2/2068/22 and 2/3116/23).

Figure captions

Fig.1. SAD pattern of the annealed NbN film deposited on the Si substrate.

Fig.2. SAD patterns of the $MgB_2$ / NbN interface.

Fig.3. MMMA losses $\delta P$ from intergranular weak links ($B_{mod}$ = 0.05 mT) and moving vortices ($B_{mod}$ = 0.02 mT) measured on the $MgB_2$ thin films.

Fig.4a,b. Magneto-optical image of the $MgB_2$ film. Bright areas represent high magnetic induction. White rectangles denote two different areas which were used for the determination of the $J_C$ (a). Transport critical current density $J_C$ at the temperature of 13.2 K for the $MgB_2$ film on the NbN buffered Si substrate as a function of magnetic field (b).

Fig.5. Measured dependendence of differential conductance (dotted line) and fitted by the BTK model curve (solid line).

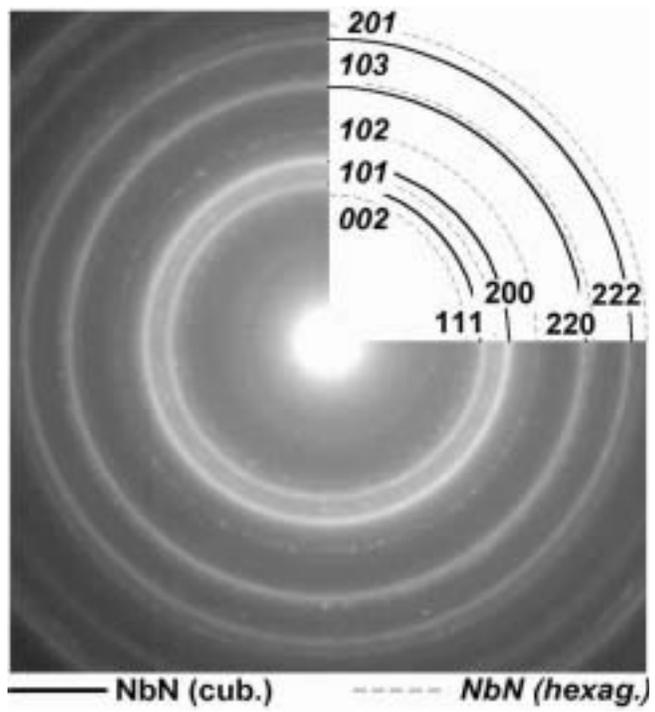

Fig. 1

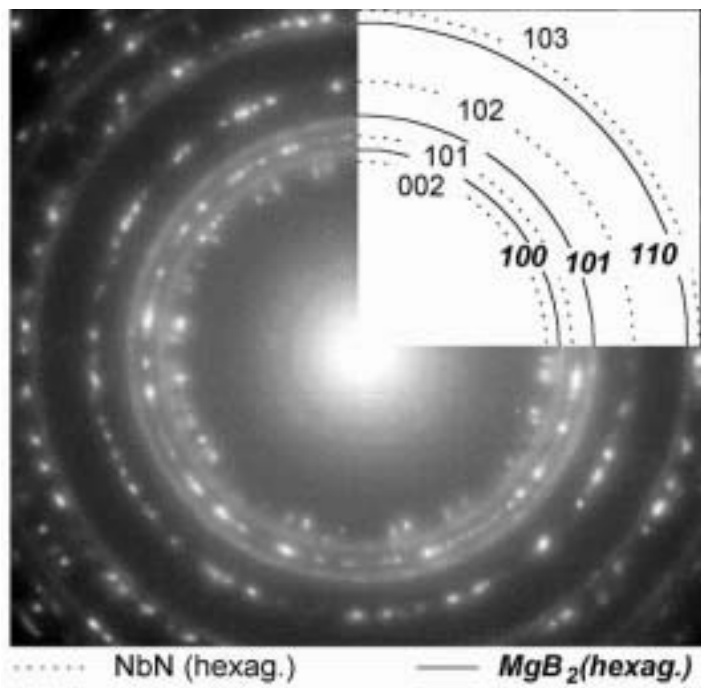

Fig. 2

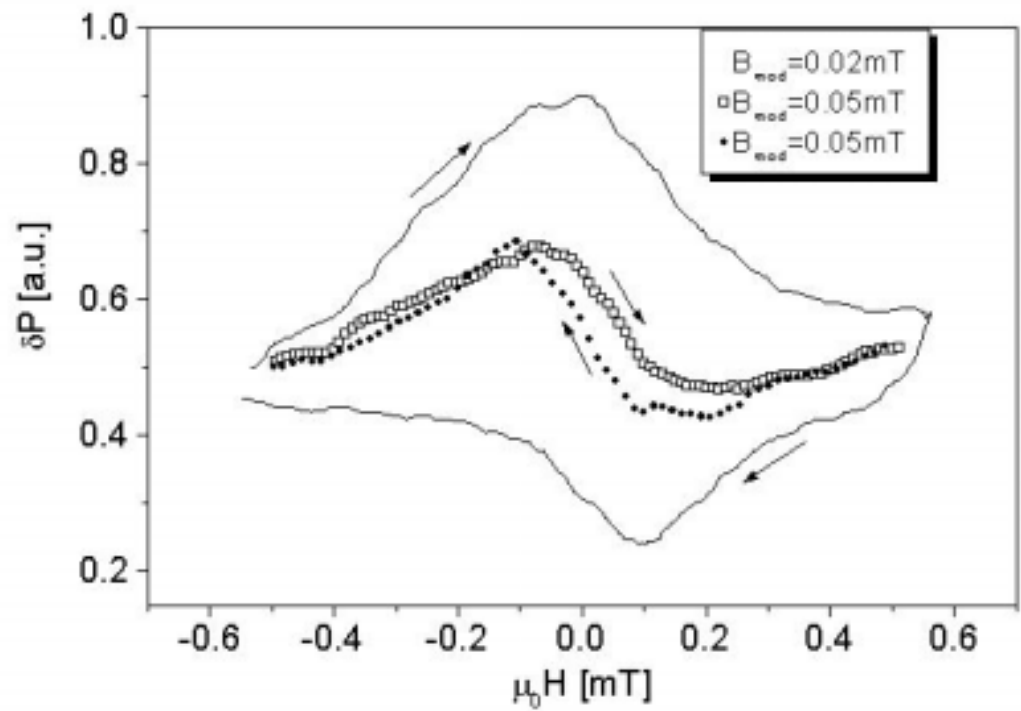

Fig. 3

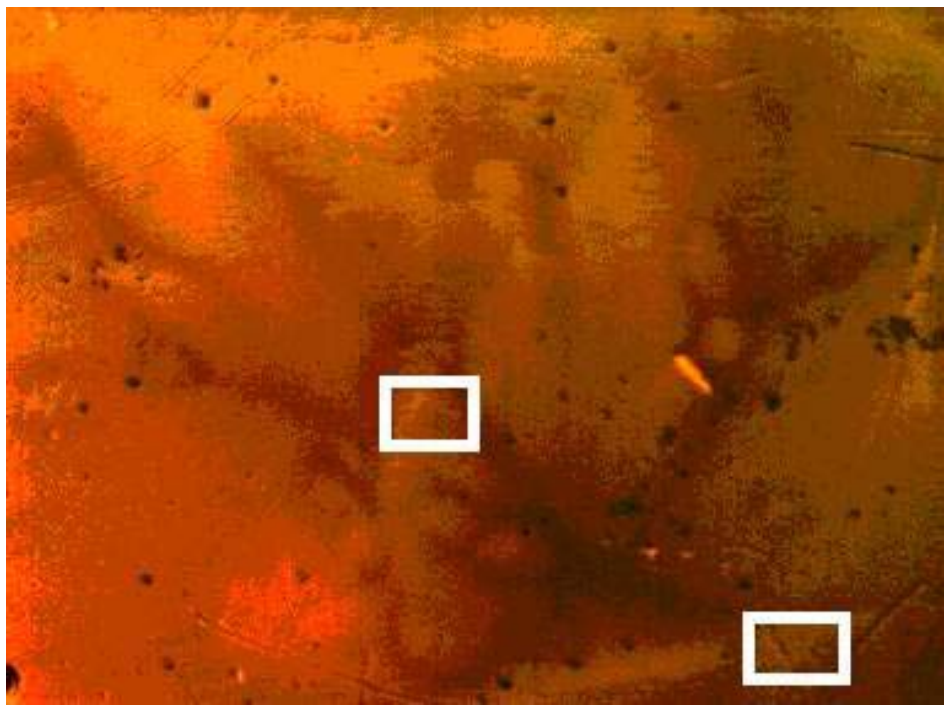

Fig. 4a

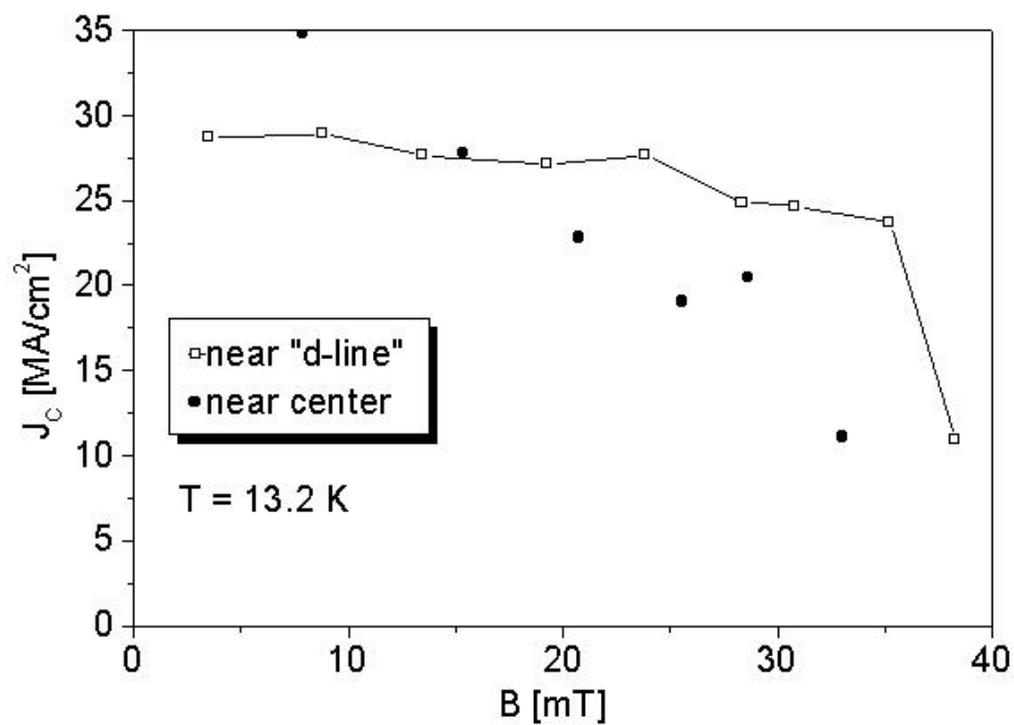

Fig, 4b

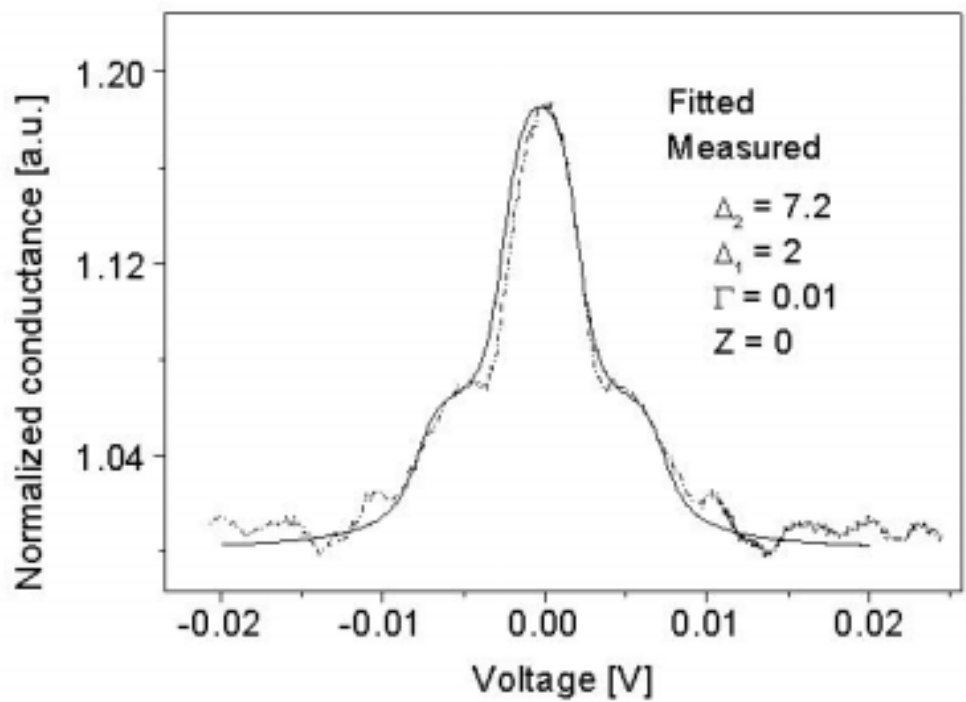

Fig. 5